# Spin singlet-triplet transition in a Si-based two-electron double quantum dot molecule

S. D. Lee, [1] S. J. Kim, [1] J. S. Kang, [1] Y. B. Cho, [1] J. B. Choi, [1,*] Sooa Park, [2] S.-R. Eric Yang, [2] S. J. Lee, [3] and T. H. Zyung[3]

[1]*Dept. of Physics & Institute for NanoScience & Technology, Chungbuk National University, Cheongju 361-763, Korea,*
[2]*Dept. of Physics, Korea University, Seoul 136-701, Korea and*
[3]*Electronics & Telecommunications Research Institute, Daejeon 305-350, Korea*

We report a successful measurement of the magnetic field-induced spin singlet-triplet transition in silicon-based coupled dot systems. Our specific experimental scheme incorporates a lateral gate-controlled Coulomb-blockaded structure in Si to meet the proposed scheme of Loss and DiVincenzo [1], and a non-equilibrium single-electron tunneling technique to probe the fine energy splitting between the spin singlet and triplet, which varies as a function of applying magnetic fields and interdot coupling constant. Our results, exhibiting the singlet-triplet crossing at a magnetic field for various interdot coupling constants, are in agreement with the theoretical predictions, and give the first experimental demonstration of the possible spin swapping occurring in the coupled double dot systems with magnetic field.

There are several proposals for scalable solid-state quantum bits (qubits). One type is charge qubits, such as superconducting boxes and excitons [2, 3]. The other type is spin qubits, whose dephasing times can be on the order of microseconds, several orders longer than those of charge qubits. Loss and DeVincenzo [1] first proposed a quantum gate mechanism based on spins of two laterally coupled QDs containing two electrons. In such a two-electron double QD the ground-state spin configuration can be either spin-singlet (S=0) or spin-triplet state (S=1), depending on the values of applied magnetic field and interdot coupling constant. The most remarkable feature is that the spin exchange J changes its sign from a positive to negative value at some magnetic field over a wide range of the interdot coupling constant. This singlet-triplet crossing at a magnetic field is an essential feature of their proposal for the quantum-gate operation since it allows making a spin swap, which can, combined with single-qubit rotations, assemble all quantum algorithms. This change of sign in J can be induced by magnetic and/or electric fields.

Despite its fundamental and practical importance, the progress in the experimental study of the spin exchange in the two coupled dots has been considerably slow as compared to extensive works for mesoscopic studies on electron transport [*4*]. This is not only because two-dimensional (2D) size fluctuations for each dot occurring during the fabrication process are inevitable, but also because it is difficult to probe the fine energy difference between spin singlet and triplet states of two electrons in the coupled QD systems. In our work we have overcome these difficulties by incorporating two specific experimental schemes for probing the spin exchange. Firstly, lateral gate-controlled Coulomb-blockade structures consisting of two coupled Si QDs are used. The gated Coulomb-blockade structure is essential since it can control both the number of electrons N and the interdot coupling constant in the double dots. The Coulomb blockade regime corresponding to N=2 must be specified, which meets the proposed scheme of Loss and Divincenzo [1]. Moreover, electrons in Si have extremely long spin lifetimes of about ms which is due to silicon's very weak spin-orbit coupling (~$10^4$ times low as compared to GaAs [5–7]). Secondly, the single-electron tunneling spectroscopy in the presence of a finite source drain voltage allows us to explore the fine structures of the coupled double dots corresponding to the excited states as well as the ground states. This non-equilibrium transport spectroscopy was already used successfully to probe the excited states of a single quantum dot [8]. Here we report a successful measurement of the magnetic field-induced spin singlet-triplet transition in silicon-based coupled dot systems. Our results, exhibiting the singlet-triplet crossing at a magnetic field for various interdot coupling constants, are in agreement with the theoretical predictions, and give the first experimental demonstration of the possible spin swapping occurring in the coupled double dot systems in the presence of a magnetic field.

The lateral gated-two coupled dots were fabricated on silicon-on-insulator (SOI) structure by pattern-dependent oxidation (PADOX) method. The SOI wafer, prepared by unibond method, consists of p-type Si substrate, 180nm-buried $SiO_2$ and 80nm-top Si. Figure 1 shows a schematic diagram of the resulting Coulomb-blockaded device structure and a scanning electron micrograph picture of the active channel which consists of two coupled QDs. The channel was first defined by e-beam lithography and followed by reactive ion etching to a narrow wire of 100nm-length and 15nm-width which abruptly widens into source and drain carrier reservoirs. Subsequent PADOX process (i) further reduces the silicon channel, (ii) generates a small quantum island with a 80nm-length and 10nm-width by oxidation-induced stress at the central part of the wire, and (iii) creates tunnel barriers at both sides in a self-aligned manner [9 – 11]. Three independent metal gates are incorporated to the Coulomb island. Biasing the middle side gate results in a stronger electrostatic effect on the current channel and produces a potential barrier at the middle point in the channel, yielding two identical dots of a size < 30x10nm each. Both end-side gates $V_{sg1}$ and $V_{sg3}$ are auxiliary gates designed to give, respectively, electrostatic contact potential barriers between the source and channel and between the drain and channel, in addition to two tunnel barriers at both sides of the Coulomb island channel already generated in a self-aligned manner by oxidation-induced stress of the PADOX process. The effect of interdot coupling on the SET current can be explored from weak to strong coupling regimes by adjusting the middle inter-gate $V_{sg2}$.

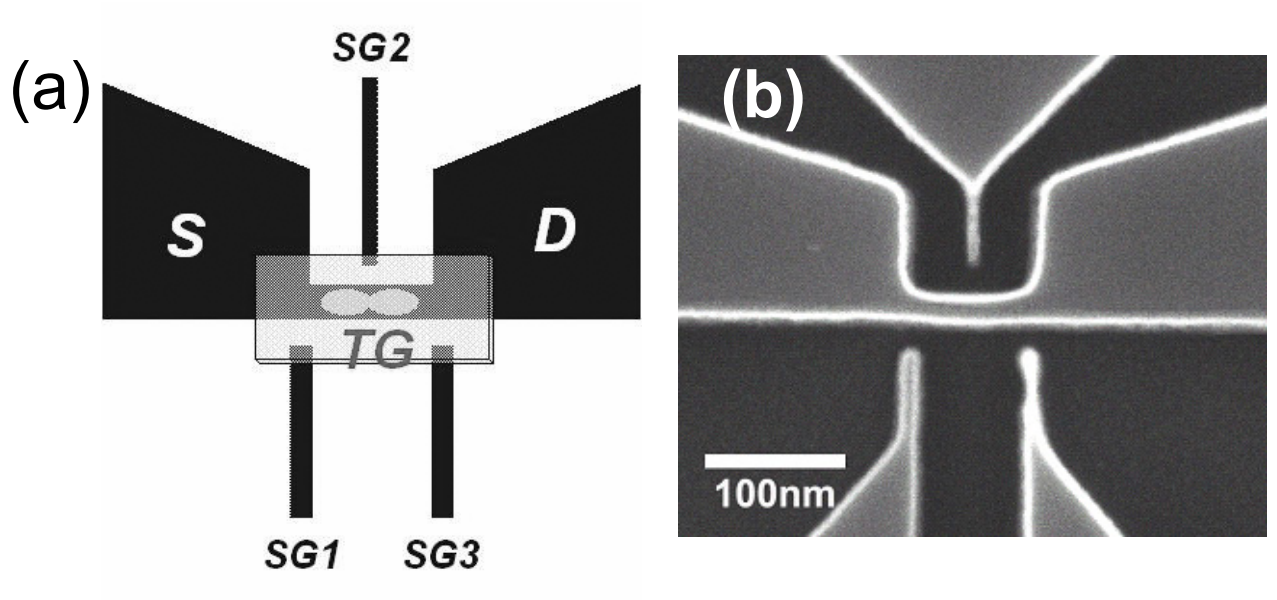

Figure 1. (a) A schematic diagram of our SET device structure. Three independent metal gates are attached to the Coulomb island. (b) A scanning electron micrograph picture of our SET device structure. Biasing the middle side gate results in a stronger electrostatic effect on the Coulomb channel and produces a potential barrier at the middle point in the channel, yielding two identical dots of a size < 30x10nm each.

Single-electron tunneling measurements were carried out in a non-equilibrium transport regime. Figure 2(a) shows typical SET currents measured as a function of top-gate voltage under a bias. Some of our samples exhibit a modulation beating in Coulomb oscillations, probably due to disorder in the dots. In contrast, samples studied here do not display such a beating, which indicates that two coupled dots are nearly identical to each other. This is mainly because two coupled dots are formed simultaneously in the same channel by the electrostatic potential barrier at the middle point. The first current peak appears at $V_g \approx 3.15$V. Its magnitude is very weak, only a few pA level as compared to other peaks, but still clearly seen for all magnetic fields. The amplitudes of the Coulomb peaks fall exponentially with decreasing Vg, and extrapolate to zero near the small peak at Vg=3.15V. This is consistent with the theory that the transmission peak amplitudes fall exponentially with electron number. Moreover, as seen in Fig. 2(b), the slope of the first Coulomb diamond becomes nearly infinite, in contrast to those of the $2^{nd}$ and $3^{rd}$ diamonds, which indicates no further Coulomb peak is present below 3.15V. Therefore, we identify the first peak as the current associated with the tunneling of the first electron into the dot system. We focus on the $2^{nd}$ current peak since it corresponds to the electron transport through the energy states of two-electron coupled QDs. This $2^{nd}$ Coulomb peak is seen to split into a set of two peaks. We point out that the higher energy peak in this $2^{nd}$ current set starts to appear when $V_{ds}$ >500μV, as seen in Fig. 2(b), which is a grey scale contour plot of $I_{ds}$ as a function of $V_{ds}$ and $V_g$. This implies that transport occurs in the non-equilibrium regime for $V_{ds}$ >500μV and the higher peak corresponds to the excited state of the two-electron QDs, while the lower peak its ground state. The x- and y-scales of the Fig. 2(b) are chosen to make a clear display of the bias-dependent spin splitting. To prevent the sample from a possible damage due to the high voltage bias we did not measured I-Vg when the bias voltage is above 10mV. However, one can already see the presence of a half diamond in the range 3.27-3.33V.

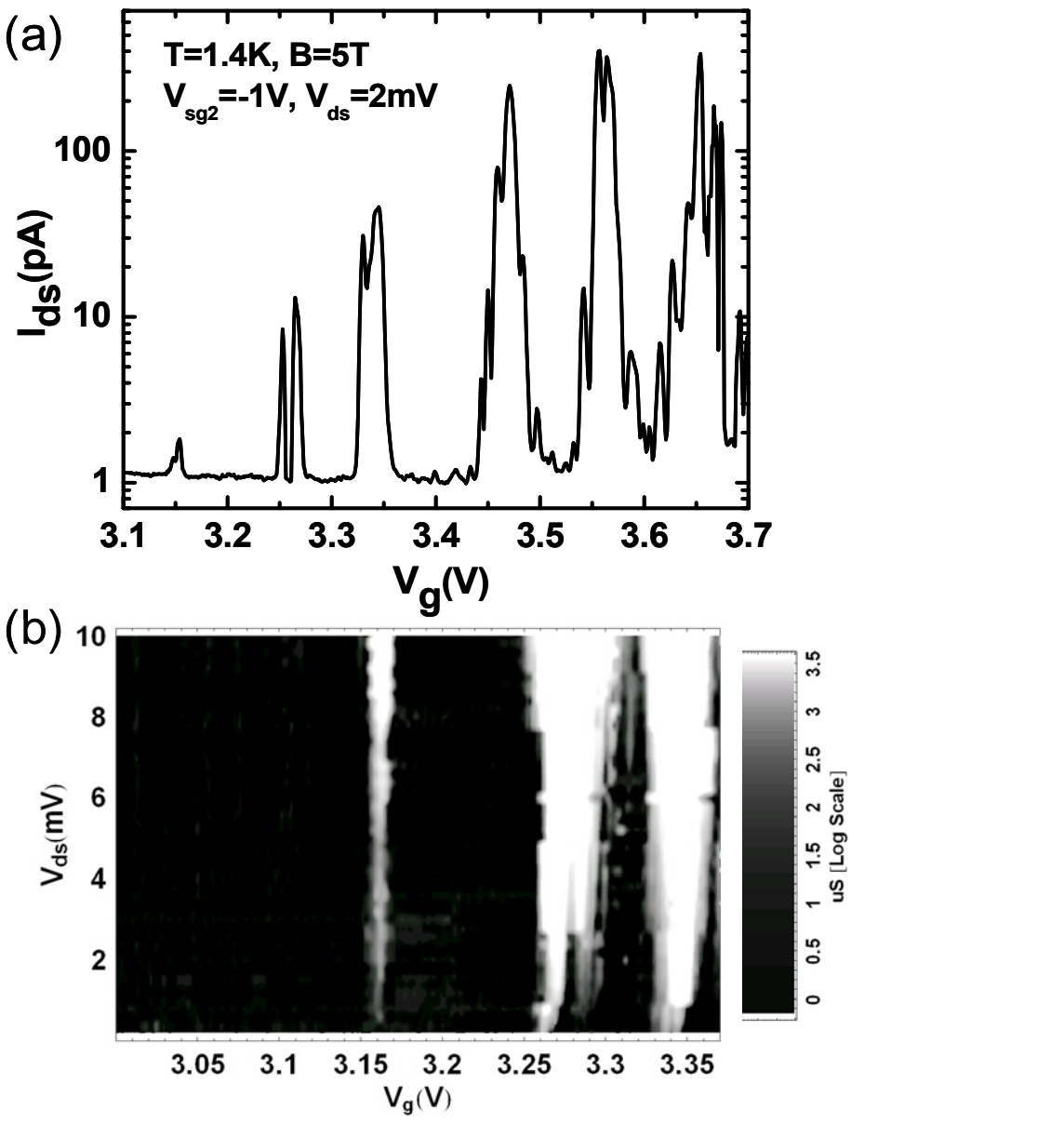

Figure 2. (a) Typical SET currents measured as a function of top-gate voltage for $V_{ds}$=2mV, T=1.4K, B=5T, and $V_{sg2}$=-1V. The $2^{nd}$ current peak, corresponding to the electron transport through an energy level of two-electron coupled QDs, is seen to split into a set of two small peaks. This $2^{nd}$ set of two peaks is found to exhibit significant magnetic field dependence, as shown in the Fig 4. (b) A grey scale contour plot of $I_{ds}$ as a function of $V_{ds}$ and $V_g$. The higher energy peak in the $2^{nd}$ current set starts to appear when $V_{ds}$ >500μV.

We explain qualitatively the feature of the $2^{nd}$ set of two peaks as follows. Fig. 3 illustrates the schematic diagram for the lateral potential energy of the two coupled QD system in a non-equilibrium regime. The ground state of the double dot is assumed to be the spin singlet state. The $2^{nd}$ electron tunnels through the singlet ground state $\varepsilon_s$ when the value of the top-gate voltage $V_g$ is such that the singlet state approaches the Fermi level of the source metal (Fig. 3(a)). This gives a rise to the $1^{st}$ peak in the $2^{nd}$ set of two peaks. When $V_g$ increases further, the singlet ground state goes down below the bottom of the energy of the source metal, and the current stops to flow since there is no density of states (DOS) in the source metal available for tunneling (Fig. 3(b)). In this regime, the current is zero, not because it is Coulomb-blockade, but because of the absence of available DOS. When $V_g$ increases even more, an excited triplet-state goes down below $\varepsilon_f$ of the source (Fig. 3(c)), and the $2^{nd}$ electron can now tunnel through the triplet state. Finally, when $V_g$ increases further and the triplet state goes down below the bottom of the energy of the source (Fig. 3(d)), the current stops again. According to this model, the currents of the $2^{nd}$ set of two peaks flow always through either singlet or triplet states of two electrons in the double dot. We remark here that this explanation is valid because the DOS of 2DEG source is small, $D_S \approx 5\times10^{10}$cm$^{-2}$, and its corresponding Fermi level $\varepsilon_f$ ≈300μeV. Moreover, the single-electron charging energy U, deduced directly from the measurement data, is estimated to be ≅ 7.5meV, which is much larger than the bias voltage drop. Therefore, the non-equilibrium transport through an excited state is expected to occur for $V_{ds} \geq J$, where J (≤1meV) is the energy split between singlet and triplet states. Note that although the system consists of two coupled dots, there is only one current peak in the equilibrium regime, which is quite different from other works on similar structures [12,13]. The reason for this is that the distance between our coupled two dots is comparable to the diameter of each dot, and, consequently, the interdot Coulomb interaction is of the same order as the intradot Coulomb interaction. The 2nd electron entering a dot will thus be blockaded by the 1st electron in the other dot, which results in only one peak even in the equilibrium transport.

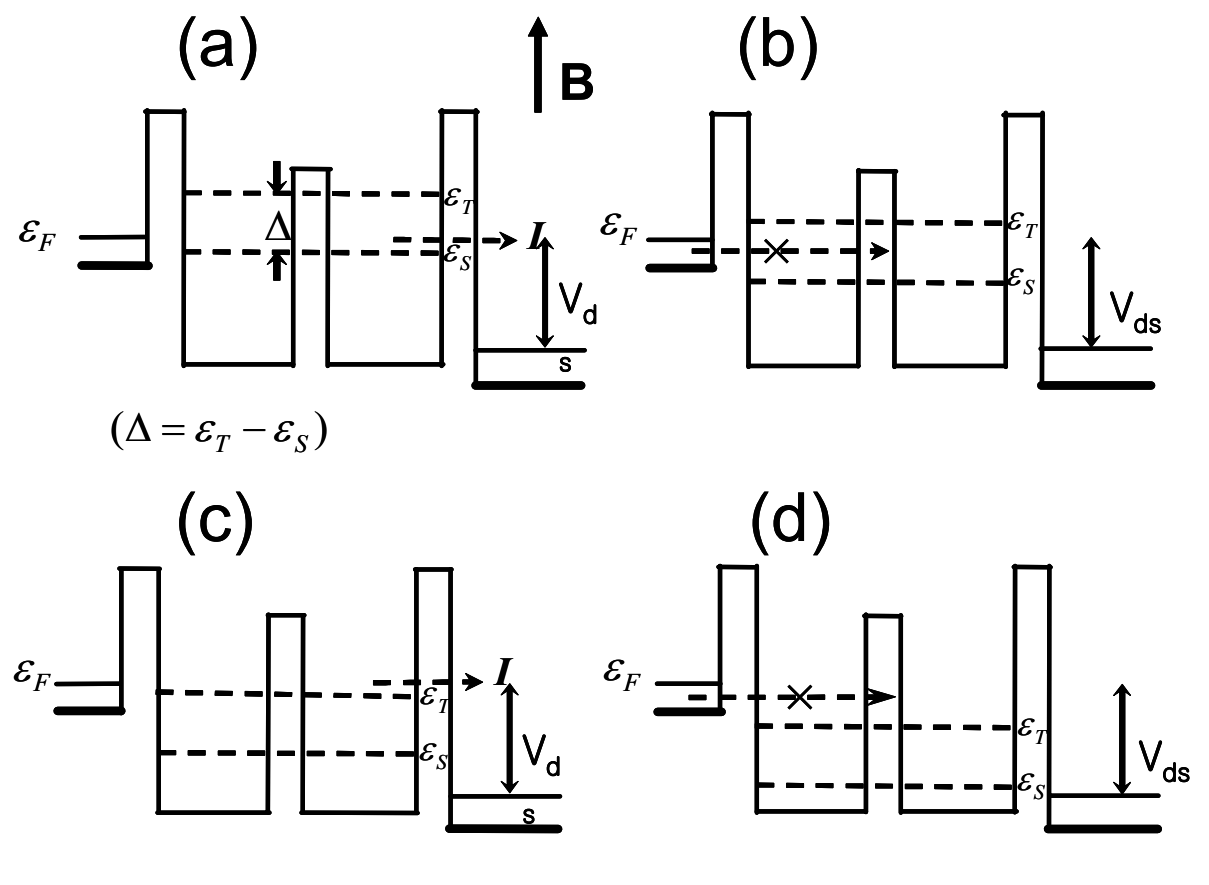

Figure 3. A schematic diagram illustrating the lateral electron potential energy of the two- coupled QD system in a non-equilibrium transport regime, where the ground state is the spin singlet. (a) There is already one electron in the system and a second electron tunnels through the singlet ground state $\varepsilon_S$, which gives rise to the 1st peak of the 2nd set of two peaks. (b) The singlet ground state energy decreases below the bottom of the energy of the source metal, and the current stops to flow since there is no density of states in the source metal available for tunneling. (c) The second electron tunnels through the triplet state. (d) The triplet state energy decreases below the bottom of the energy of the source, and the current stops again.

The most remarkable feature of the 2nd set of two peaks is found in their magnetic field dependence. Fig. 4 shows the magnetic field dependence of the 2nd set of two current peaks measured at 1.5K. As magnetic field increases the split between the two peaks decreases first, approaching almost zero at B≈3T. For B>3T, it increases again and finally decreases to zero. This magnetic field-dependence of the split between the two peaks in the 2nd set is observed for other interdot coupling constants. The Zeeman splitting of the triplet states appears not to be resolved clearly in our experiment since the broadening of peaks is comparable to the energy splitting between the states. Fig. 5 summarizes the split of the 2nd set of two peaks plotted as a function of magnetic field for three different interdot coupling constants. The y-axis of the main figure displays the values of J given in units of mV (Note that, in contrast, the scale of the y-axis in the inset is in meV). These values can be converted into meV by using a converting factor, α, which is defined by $\alpha = \frac{C_g}{C_{total}} = \frac{C_g}{C_{self} + C_g} = 0.13$, where the gate capacitance $C_g$ is 1.4aF and the self capacitance of the dot $C_{self}$ is 9.2aF. The singlet-triplet transition field, where J=0, is observed to shift slightly to high fields with increasing interdot coupling. We attribute this behavior to the spin exchange of the two-electron coupled dots, which varies with the applied magnetic field. Note that the singlet-triplet crossing, occurring over a wide range value of interdot coupling constants, is caused by the long-range Coulomb interaction. Theoretical calculations by Burkard et al. (*14*) and others [15, 16] show that the spin exchange changes its sign from positive to negative at some magnetic field..

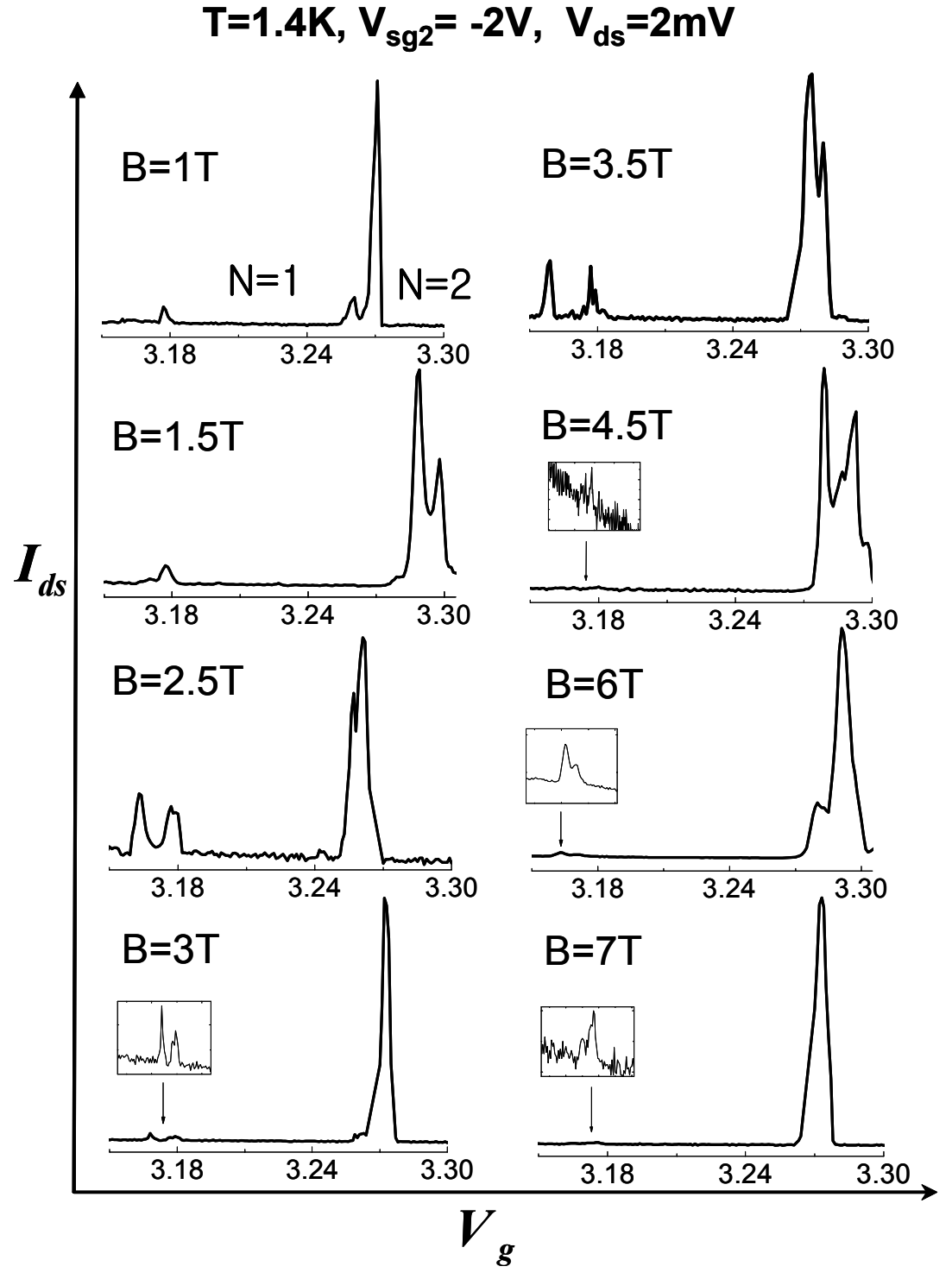

Figure 4. $I_{ds}$ vs. $V_g$ for different values of magnetic fields measured at 1.5K. As magnetic field increases the split between two peaks becomes decreasing first, approaching almost zero at B≈3T. For B>3T, it increases again and finally decreases to zero.

Burkard et al have calculated the exchange energy J of two electrons in coupled dots defined by the potential, $V(x,y) = \frac{m\omega^2}{2}\left[\frac{1}{4a^2}(x^2 - a^2)^2 + y^2\right] + exE$, where $2a$ is the interdot distance, $\omega$ is the characteristic frequency of each quantum dot potential, m is the effective mass, and $E$ is the electric field between the source and drain. We calculate the energy difference between the lowest energy the triplet state and the singlet state using the expression for J obtained by Burkard et al. The inset in Fig. 5 displays this energy difference J for three different values of $a$. We have used the following parameters: the characteristic frequency of the quantum dot $\omega$ =1.5meV, Si effective mass of 0.2m$_0$, the interdot distance $2a$ =19.6, 19.8, 20nm, the Zeeman splitting $g\mu_B B = 0.116B$[meV], and the voltage

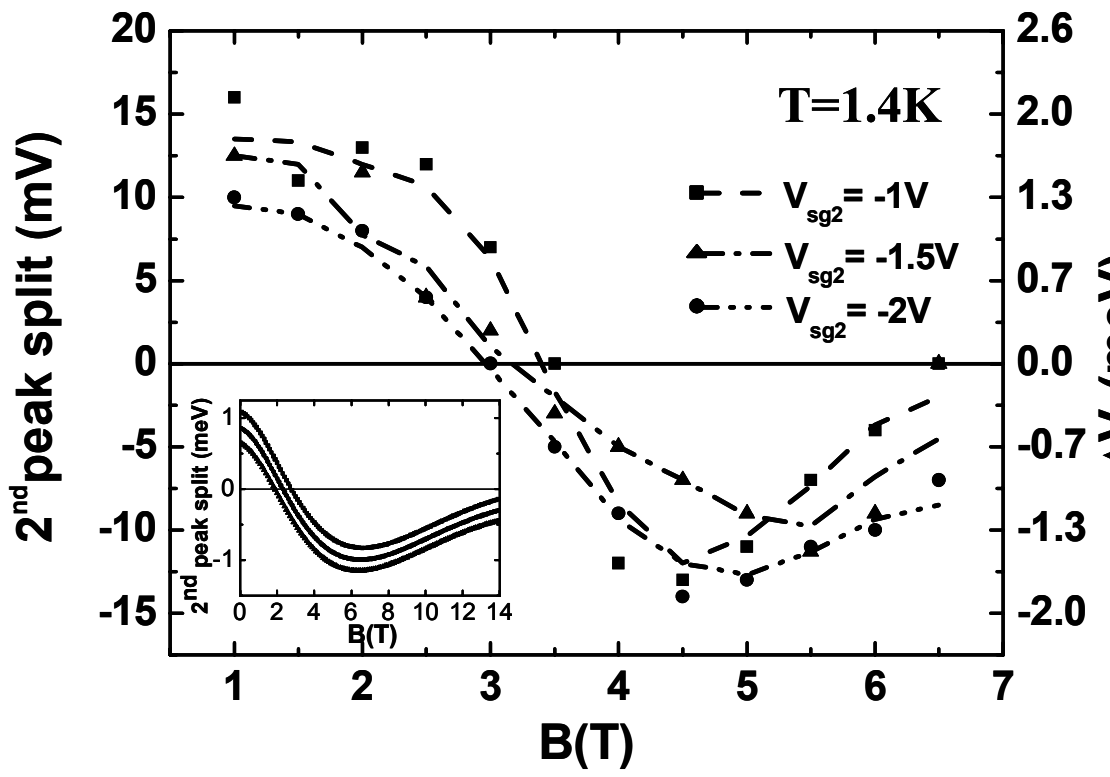

Figure 5. Split in the 2nd set of two peaks plotted as a function of magnetic field for three different interdot coupling constants. Solid lines drawn along the measured data are polynomial data fits. Inset shows theoretical results obtained by using an expression of Burkard et al (15) for the spin exchange between two electrons in coupled dots for three different interdot distances

drop between the dots $eEa$ =3.5meV. Note that overall features of the magnetic field dependence are in qualitative agreement with the measured values. (The experimental data can be also fitted using a slightly different model which contains the barrier height as an input parameter [15]. In this model the values of J for different barrier heights will cross each other at some magnetic field values, in agreement with the experimental data. However, in this model the absolute magnitude of J is always the largest for the smallest barrier height, in disagreement with the experimental data for high-field regime of B>5T). From these theoretical considerations we estimate that J is of order 1meV. Figure 6 shows the inter-gate voltage dependence of the split of two peaks at a fixed magnetic field. As illustrated in the inset, for three different magnetic fields the splits decay exponentially as a function of the interdot coupling, which is also consistent with the theoretical prediction on the exchange coupling for large interdot distance [14, 15]. This agreement implies that adjusting the central potential barrier and/or interdot separation by the inter-gate voltage can give an efficient control of the splitting between the singlet and triplet states.

The vanishing of J at a magnetic field can be exploited for spin swapping and for the implementation of two-qubit gate. A constant uniform magnetic field B≈3T can be applied to the two coupled QDs to tune J close to zero, and following this a small gate pulse or a small local magnetic field can be applied for switching J on and off. Note that our coupled two-dot system was fabricated on silicon wafer. Electrons in silicon are very promising for spintronics and quantum information processing since they have extremely long spin lifetimes of about ms which is due to silicon's very weak spin-orbit coupling. Moreover, the silicon VLSI technology is expected to accelerate our progress towards solid-state implementations of the scalable quantum computer in Si.

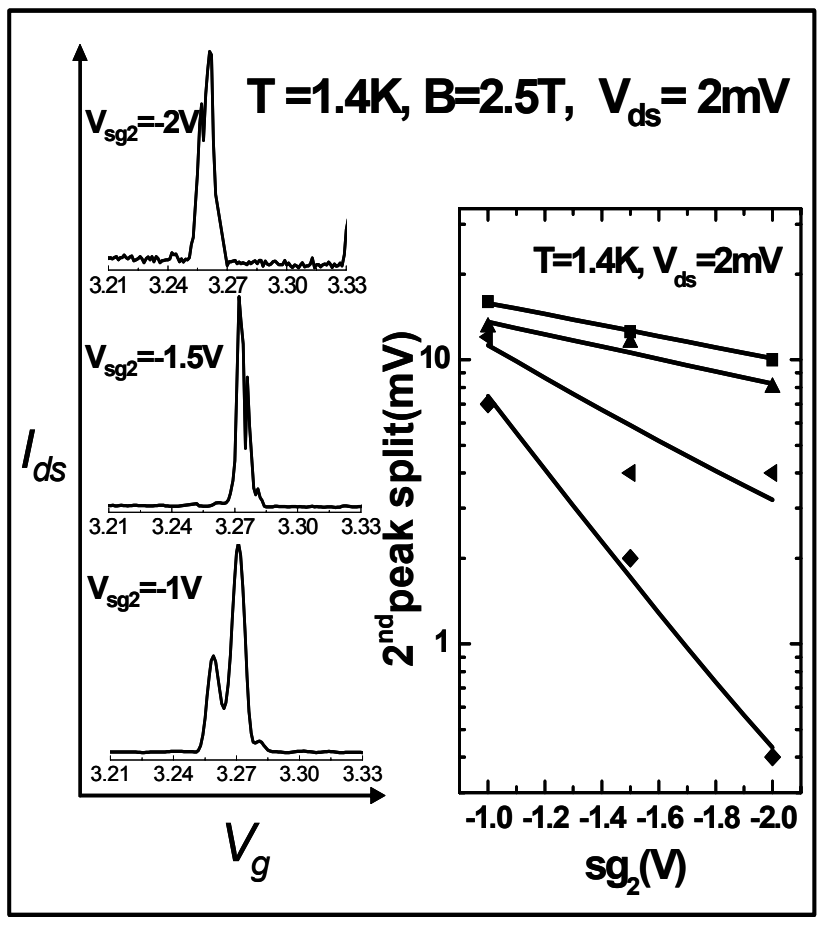

Figure 6. Inter-gate voltage dependence of the split of two peaks at a fixed magnetic field for T=1.4K. As illustrated in the inset, for three different magnetic field the splits decay exponentially for small interdot coupling, which is consistent with the theoretical prediction for the exchange coupling.

We acknowledge useful discussions with G. Burkard , D.P. Divincenzo , D. Loss, and R. Nieminen. This work was supported by Korea Ministry of Science & Technology through the Frontier 21 National Program for Tera-level Nanodevices (TND) and in part by Korea Science & Engineering Foundation through Quantum-functional Semiconductor Research Center (QSRC).

* Electronic address: jungchoi@chungbuk.ac.kr